\documentclass[prd,preprintnumbers,twocolumn,eqsecnum,floatfix,a4paper,nofootinbib,superscriptaddress]{revtex4-1}

\usepackage{color}
\usepackage{amsmath,amssymb,graphicx}
\usepackage{bm,lipsum}
\usepackage{times}
\usepackage{microtype}
\usepackage[utf8]{inputenc}
\usepackage{booktabs}
\usepackage{subfigure}
\usepackage[normalem]{ulem}
\usepackage[varg]{txfonts}
\usepackage{bm,graphics,graphicx,epsfig,color,ulem}
\usepackage[colorlinks, pdfborder={0 0 0}]{hyperref}
\definecolor{LinkColor}{rgb}{0.75, 0, 0}
\definecolor{CiteColor}{rgb}{0, 0.5, 0.5}
\definecolor{UrlColor}{rgb}{0, 0, 0.75}
\hypersetup{linkcolor=LinkColor}
\hypersetup{citecolor=CiteColor}
\hypersetup{urlcolor=UrlColor}
\allowdisplaybreaks
\def\linechange{\nonumber\\&&}

\begin{document}


\title{Testing the multipole structure of compact binaries using
gravitational wave observations}
\author{Shilpa Kastha}\email{shilpakastha@imsc.res.in}
\affiliation{Institute of Mathematical Sciences, HBNI, CIT Campus, Chennai-600113, India}
\author{Anuradha Gupta} \email{axg645@psu.edu}
\affiliation{Institute for Gravitation and the Cosmos, Department of Physics, Penn State University, University Park PA 16802, USA}
\author{K. G. Arun} \email{kgarun@cmi.ac.in} 
\affiliation{Chennai Mathematical Institute, Siruseri, 603103, India}
\affiliation{Institute for Gravitation and the Cosmos, Department of Physics, Penn State University, University Park PA 16802, USA}
\author{B. S. Sathyaprakash} \email{bss25@psu.edu} 
\affiliation{Institute for Gravitation and the Cosmos, Department of Physics, Penn State University, University Park PA 16802, USA}
\affiliation{Department of Astronomy and Astrophysics, Penn State University, University Park PA 16802, USA}
\affiliation{School of Physics and Astronomy, Cardiff University, Cardiff, CF24 3AA, United Kingdom}
\author{Chris Van Den Broeck}\email{vdbroeck@nikhef.nl}
\affiliation{Nikhef - National Institute for Subatomic Physics, Science Park 105, 1098 XG Amsterdam, The Netherlands}
\affiliation{Van Swinderen Institute for Particle Physics and Gravity, University of Groningen, Nijenborgh 4, 9747 AG Groningen, The Netherlands}
\date{\today}

\begin{abstract}
We propose a novel method to test the consistency of the multipole
moments of compact binary systems with the predictions of general
relativity (GR). The multipole moments of a compact binary
system, known in terms of symmetric and trace-free tensors, are used to calculate the gravitational waveforms from compact binaries
within the post-Newtonian (PN) formalism. For nonspinning compact
binaries, we derive
the gravitational wave phasing formula, in the frequency domain,
parametrizing each PN order term in terms of the multipole moments which
contribute to that order. 
Using GW observations, this {\it{parametrized multipolar phasing}} would allow us to derive the bounds on possible departures from the multipole
structure of GR and hence constrain the parameter space of
alternative theories of gravity. We compute the projected accuracies with which 
the second-generation ground-based detectors, such as the Advanced Laser Interferometer 
Gravitational-wave Observatory (LIGO), the third-generation detectors such as the Einstein Telescope and Cosmic Explorer, as well as the space-based detector Laser Interferometer Space Antenna (LISA) will be able to 
measure these multipole parameters. We find that while Advanced LIGO can measure 
the first two or three multipole coefficients with good accuracy, Cosmic Explorer and the Einstein Telescope may be able to measure the first four
multipole coefficients which enter the phasing formula. Intermediate-mass-ratio inspirals, with mass ratios of several tens, in the frequency band of the
planned space-based LISA mission should be able to measure all seven
multipole coefficients which appear in the 3.5PN phasing formula. Our
finding
highlights the importance of this class of sources for probing
the strong-field gravity regime. The proposed test will facilitate the first probe of
the multipolar structure of Einstein's general relativity.
\end{abstract}
\maketitle

\section{Introduction} 
The discovery of binary black holes \cite{Discovery,GW151226,GW170104,GW170814} and binary neutron stars \cite{GW170817}
by Advanced LIGO~\cite{aLIGO} and Advanced Virgo~\cite{aVirgo} have been ground breaking for several
reasons. Among the most important aspects of these discoveries is
the unprecedented opportunity they have provided to study the
behavior of gravity in the highly nonlinear and dynamical regime
associated with the merger of two black holes (BHs) or two neutron
stars (see Refs.\ \cite{SathyaSchutzLivRev09,YunesSiemens2013} for
reviews).
The gravitational wave (GW) observations have put stringent constraints on the allowed
parameter space of alternative theories of gravity by different methods~\cite{TOG, O1BBH, GW170104}.
They include the parametrized tests of post-Newtonian
theory~\cite{AIQS06a,AIQS06b,YunesPretorius09,MAIS10,TIGER, Li:2011cg,Meidam:2017dgf}, bounding the mass of the
putative graviton and dispersion of GWs~\cite{Will98,MYW11}, testing consistency between the inspiral and
ringdown regimes of the coalescence~\cite{IMRConsistency} and the time delay between the GW
and electromagnetic signals~\cite{GW-GRB170817}. Furthermore, the bounds obtained
from these tests have been translated into bounds on the free parameters
of certain specific theories of gravity~\cite{YYP2016}. 

With improved sensitivities of Advanced LIGO and Virgo in the upcoming observing runs,
the development of third-generation detectors such as the Einstein Telescope (ET) 
\cite{ArunET09}  and Cosmic Explorer (CE) \cite{CE_WB} and the approval of funding
for the space-based mission LISA \cite{LISA}, the field of gravitational astronomy promises to deliver exciting science returns.  In addition to stellar-mass 
compact binaries, future ground-based detectors, such as ET and CE, can detect intermediate-mass black holes with a total mass of  several hundreds of solar 
masses. Such observations will not only confirm the existence of BHs in this mass 
range (see Refs.~\cite{MillerIMBHreview04,IMBHreview} for reviews), but also facilitate 
several new probes of fundamental physics via studying their
dynamics~\cite{Fregeau:2006yz,GraffIMBH2015,BrownIMRI07,Chamberlain:2017fjl}. Some of the most prominent, among these, are those using intermediate-mass-ratio inspirals, which will last longer (compared to the equal-mass binaries), and hence are an
accurate probe of the compact binary dynamics and the BH nature
of the central compact object~\cite{Ryan97,BrownIMRI07}. 

The space-based LISA mission, on the other hand, will be sensitive to millihertz GWs
produced by the inspirals and merger of supermassive BH binaries in the
mass range $\sim 10^4$-$10^7 M_{\odot}$. These sources may also have a
large diversity in their mass ratios ranging from comparable mass
(mass ratio $\lesssim10$) and 
intermediate mass ratios (mass ratio $\gtrsim 100$) to extreme mass ratios
(mass ratio $\gtrsim 10^6$) where a stellar mass BH spirals into the central supermassive BH with several millions of solar
masses~\cite{ThorneEMRI95,SchutzEMRI96}. This diversity together with
the sensitivity in the low-frequency window makes LISA a
very efficient probe of possible deviations from general relativity (GR) in different regimes
of dynamics~(see Refs.~\cite{GairLivRev,BertiCQGRev,SathyaSchutzLivRev09,APrev13} for reviews).
 
Setting stringent limits on possible departures
from GR as well as constraining the parameter space of exotic compact
objects that can mimic the properties of
BHs~\cite{Giudice2016,Chirenti2016,Cardoso2016,Cardoso2017,Johnson-McDaniel2018,KAM2017,Dhanpal2018,Krishnendu2018}, are among the principle
science goals of the next-generation detectors. They should also be
able to detect any new physics, or modifications to GR, if present. 

Formulating new methods to carry out such tests is crucial in order to
efficiently extract the physics from GW observations. The dynamics of a
compact binary system is conventionally divided into the {\it adiabatic
inspiral, rapid merger} and {\em fast ringdown} phases. During the inspiral phase
the orbital time scale is much smaller than the radiation backreaction time scale. The post-Newtonian (PN) approximation to GR has proved
to be a very effective method to describe the inspiral phase of a compact binary of comparable masses~\cite{Bliving}.  A description of the  highly nonlinear 
phase of the merger of two compact objects needs numerical solutions to Einstein's
equations~\cite{Pretorius07Review}. The ringdown radiation of GWs by the
merger remnant, can be modeled within the framework of BH
perturbation theory~\cite{TSLivRev03}. In alternative theories of
gravity, the dynamics of the compact binary during these phases of
evolution could be quite different from that predicted by GR.  Hence
observing GWs is the best way to probe the presence of non-GR physics
associated with this phenomenon.

One of the most generic tests of the binary dynamics has been the measurement of the PN
coefficients of the GW phasing
formula~\cite{BSat94,BSat95,AIQS06a,AIQS06b,YunesPretorius09,MAIS10,TIGER}. This test
captures a possible departure from GR by measuring the PN 
coefficients in the phase evolution of the GW signal. In addition to the source
physics, the different PN terms
in the phase evolution contain information about different nonlinear interactions
the wave undergoes as it propagates from the source to the detector.
Hence the predictions for these effects in an alternative theory of
gravity could be very different from that of GR, which is what is being
tested using the parametrized tests of PN theory.

In this work, we go one step further and propose a novel way to test the
multipolar structure of the gravitational field of a compact binary
as it evolves through the adiabatic inspiral phase. The multipole moments of the compact binary (and interactions between them),  are responsible for the various physical
effects we see at different PN orders. By measuring these
effects we can constrain the multipolar structure of the
system. The GW phase and frequency evolution is obtained from the energy flux of GWs
and the conserved orbital energy by using the energy balance argument, which
equates the GW energy flux ${\cal F}$ to the decrease in the binding energy 
$E_{\rm orb}$ of the binary~\cite{BDI95}
\begin{equation}
{\cal F}=-\frac{d}{dt} E_{\rm orb}.
\end{equation}

In an alternative theory of gravity, one or more multipole moments
of a binary system may be different from those of GR.  For instance, in
Ref.\,\cite{Endlich:2017tqa}, the authors discuss how effective-field-theory-based 
approach can be used to go beyond Einstein's gravity by introducing
additional terms to the GR Lagrangian which are higher-order operators constructed out of the Riemann tensor, but suppressed by appropriate scales comparable to the curvature of  the compact binaries.
They find that such generic
modifications will lead to multipole moments of compact binaries
that are different from GR. Our proposed method aims to constrain such
generic extensions of GR by directly measuring the multipole moments of the compact
binaries through GW observations. 

In this work, we assume that the conserved orbital energy of the binary is the
same as in GR and modify the gravitational wave flux by
deforming  the multipole moments
which contribute to it by employing the multipolar post-Minkowskian
formalism~\cite{Bliving,BDI95}. We then rederive the GW phase and its
frequency evolution (sometimes referred to as the {\em phasing formula}) explicitly 
in terms of the various deformed multipole moments (In the Appendix we provide a more general
expression for the phasing where the conserved energy is also deformed
at different PN orders, in addition to the multipole moments of the
source.). We use this parametrized 
multipolar phasing formula to measure possible deviations from GR and discuss 
the level of bounds we can expect from the current and next-generation ground-based 
GW detectors, as well as the space-based LISA detector.  We obtain the measurement 
accuracy of the system's physical parameters and the deformation of the multipole moments using 
the semianalytical Fisher information matrix~\cite{Rao45,Cramer46}. These results
are validated for several configurations of the binary system by Markov chain Monte 
Carlo (MCMC) sampling of the likelihood function using the {\tt emcee}~\cite{emcee} 
algorithm. 

We find that Advanced LIGO-like detectors can constrain at most
two of the leading multipoles, while a third-generation detector, such as
ET or CE, can set constraints on as many as four of the leading
multipoles. The space-based LISA detector will have the ability to set
good limits on  all seven multipole moments that contribute to the 3.5PN phasing
formula, making it a very accurate probe of the highly nonlinear
dynamics of compact binaries. 

The organization of the paper is as follows. In Sec.~\ref{PMGWphasing}
we describe the  basic formalism to obtain the parametrized multipolar
GW phasing formula. In Sec.~\ref{Param_est} we briefly
explain  the two parameter estimation schemes (Fisher information matrix
and Bayesian inference) used in our analysis, followed
by Sec.~\ref{results} where we discuss the results we obtain for various
ground-based and space-based detectors. Section~\ref{sec:conclusion}
summarizes the paper and lists some of the follow-ups we are pursuing.

\section{Parametrized multipolar gravitational wave phasing}\label{PMGWphasing}
The two-body problem in GR can
be solved perturbatively using PN theory in the
adiabatic regime, where the orbital time scale is much smaller than the
radiation backreaction time scale (see Ref.\,\cite{Bliving} for a review). 
The PN theory has given us several useful insights about various facets of
the two-body dynamics and the resulting gravitational radiation.

In the multipolar post-Minkowskian (MPM)
formalism~\cite{Th80,BD84,BD86,B87,BD88,BD89,BD92,B95,BDI95,BIJ02,DJSdim,BDEI04}, the
important quantities such as the gravitational waveform, energy and
angular momentum fluxes  can be expressed using a combination of
post-Minkowskian approximation
(expansion in powers of $G$, Newton's gravitational constant, valid
throughout the spacetime for weakly gravitating sources), 
PN expansions (an expansion in ${1/c}$ that is valid for slowly
moving and weakly gravitating sources and applicable in the near zone of the source)
and the multipole expansion of the gravitational field valid over the entire region exterior to the source. The coefficients of post-Minkowskian expansion and
the multipole moments of the source can be further expanded as a PN
series. The multipole expansion of the gravitational field plays a central role in the analytical treatment
of the two-body problem  as it significantly helps to
handle the nonlinearities of Einstein's equations. 

The MPM formalism relates the radiation content in the far zone (at the
detector) to the stress-energy tensor of the source. The quantities in the
far zone are described by mass- and current-type {\it radiative} multipole
moments $\{U_L,V_L\}$ whereas the properties of the source are completely
described by the mass- and current-type source multipole moments $\{I_L,
J_L\}$ and the four gauge moments $\{W_L, X_L, Y_L, Z_L\}$ all of which  are the moments of the relativistic mass and current densities expressed as functionals of the stress-energy pseudo-tensor of
the source and gravitational fields. However, in GR, there is further gauge freedom to
reduce this set of six source moments to a set of two ``canonical" multipole
moments $\{M_L, S_L\}$. 
The relations
connecting these two sets of multipole moments can be found in Eqs. (97)
and (98) of Ref.~\cite{Bliving}. Furthermore, the mass- and current-type radiative multipole moments $\{U_L,V_L\}$ admit closed-form expressions in terms of $\{M_L,S_L\}$.

The source and the canonical multipole moments  are usually expressed using the
basis of 
symmetric trace-free tensors~\cite{DI91b}. The relationships between the radiative-
and source-type multipole moments incorporate the various nonlinear
interactions between the various multipoles, such as
tails~\cite{BD87,BS93,BDI95}, tails of tails~\cite{B98tail},
tail square~\cite{B98quad}, memory~\cite{Chr91,Th92,ABIQ04,Favata08}, $\ldots,$
as the wave propagates from the source to the detector (see Ref.\,\cite{Bliving} 
for more details).  

For quasicircular inspirals, the PN expressions for the orbital energy and the energy flux, together with the
energy balance argument is used in the computation of the GW phasing
formula at any PN order~\cite{BDI95,BDIWW95,BFIJ02,BDEI04}.  The PN terms in the phasing formula, hence, explicitly encode the
information about the multipolar structure of the gravitational field
of the two-body dynamics.

In this work, we separately keep track of
the contributions from various radiative multipole moments to the GW flux allowing us to
derive a parametrized multipolar gravitational wave flux and phasing formula, 
thereby permitting tests of the multipolar structure of the PN approximation to GR.
We first rederive the phasing formula for nonspinning
compact binaries moving in quasicircular orbits up to 3.5 PN order. The
computation is described in the next section. Before we proceed, we clarify that in our notation the first post-Newtonian (1PN) correction would refer to corrections of order
$v^2/c^2$, where $v=(\pi m f)^{1/3}$ is the characteristic orbital velocity of the
binary, $m$ is the total mass of the binary and $f$ is the orbital frequency.

\subsection{ The multipolar structure of the energy flux}
 
The multipole expansion of the energy flux within the MPM formalism schematically
reads as~\cite{Th80,BDI95}
\begin{equation}
{\cal F}= \sum_l
\left[\frac{\alpha_l}{c^{l-2}}U_L^{(1)}U_L^{(1)}+\frac{\beta_l}{c^{l}}V_L^{(1)}V_L^{(1)}\right],\label{eq:F}
\end{equation}
where $\alpha_l, \beta_l$ are known real numbers and $U_L,V_L$ are mass- and
current-type radiative multipole moments with $l$ indices; the superscript $(1)$
denotes the first time derivative of the multipoles.  The $U_L$ and $V_L$ can be
rewritten in terms of the source multipole moments as 
\begin{eqnarray}
U_L=M_L^{(l)}+ {\rm Nonlinear\;interaction\;terms},\\
V_L=S_L^{(l)}+ {\rm Nonlinear\;interaction\;terms},\label{eq:UVtoIJ}
\end{eqnarray}
where the right-hand side involves $l^{\rm th}$ time derivative of the mass-
and current-type source multipole moments and nonlinear interactions
between the various multipoles due to the propagation of the wave in
the curved spacetime of the source. (see Refs.~\cite{BD92,B98quad,B98tail,BIJ02}
for details). The various types of interactions can be decomposed as
follows~\cite{BDI95,BIJ02} 
\begin{equation}
\mathcal{F}=\mathcal{F}_{\mathrm{inst}}+\mathcal{F}_{\mathrm{tail}}+\mathcal{F}_{\mathrm{tail}^2}+\mathcal{F}_{\mathrm{tail}(\mathrm{tail})}\label{EFparts}.
\end{equation}
As opposed to $\mathcal{F}_{\mathrm{inst}}$ (a contribution that
depends on the dynamics of the binary at the purely retarded instant of
time, referred to as instantaneous terms),
the last three contributions $\mathcal{F}_{\mathrm{tail}}$,
$\mathcal{F}_{\mathrm{tail}^2}$ and $\mathcal{F}_{\mathrm{tail}(\mathrm{tail})}$ contain nonlinear multipolar interactions in the flux~\cite{B98tail} that depend on the dynamical history of the system and are referred to as {\it hereditary} contributions. 

In an alternative theory of gravity, the multipole moments may not be
the same as in GR; if the mass- and current-type radiative multipole moments deviate from 
their GR values by a fractional amount $\delta U_L$ and $\delta V_L,$ i.e.,
$U_L\rightarrow U_L^{\rm GR}+\delta U_L$ and $V_L\rightarrow V_L^{\rm
GR}+\delta V_L$, then we can parametrize such deviations in the
multipoles by considering the scalings 
\begin{eqnarray}
\label{scaled-moments}
U_L\rightarrow \mu_l\,U_L,\nonumber\\
V_L \rightarrow \epsilon_l \, V_L,
\end{eqnarray}
where the parameters $\mu_l=1+\delta U_L/U_L^{\rm GR}$ and  $\epsilon_l=1+\delta
V_L/V_L^{\rm GR}$ are equal to unity in GR.

We first recompute the GW flux from nonspinning
binaries moving in quasicircular orbit up to 3.5PN order with the above scaling using
the prescription outlined in Refs.~\cite{BDI95,B95,BIJ02,BDEI04}. 
With the parametrizations introduced above, the computation of the
energy flux would proceed similarly to that in GR but contributions from
every radiative multipole are now separately kept track of.

In order to calculate the fluxes up to the required PN order, we
need to compute the time derivatives of the multipole moments as
can be seen from Eqs. (\ref{eq:F}-(\ref{eq:UVtoIJ}). These are
computed by using the equations
of motion of the compact binary for quasicircular orbits given
by~\cite{BI03CM,BIJ02}
\begin{equation}
\frac{d {\bf v}}{dt}=-\omega^2\, {\bf x},
\end{equation}
where the expression for $\omega$, the angular frequency of the binary,
up to 3PN order is given by~\cite{DJSdim,BDE04,DJSequiv,ABF01,BI03CM,itoh1,BFIJ02}
\begin{eqnarray}\label{eq:EOM}
\omega^2 &=& {Gm\over r^3} \Biggl\{1+\left[-3+\nu\right]\gamma +
\left[6+{41\over 4} \nu + \nu^2 \right] \gamma^2 \nonumber \\ 
&+&\left[-10+\left(22 \ln \left( r \over {r'}_0\right)+{41\pi^2\over
64}-{75707\over 840}\right)\nu\right.\nonumber\\
&+&\left.{19\over 2}\nu^2
+\nu^3\right] \gamma^3 +{\cal O}(\gamma^4)\Biggr\},
\end{eqnarray} 
where $\gamma=G m/r c^2$ is a PN parameter, $r_0'$ is a gauge-dependent length scale which does not appear when observables, such as
the energy flux, are expressed in terms of gauge-independent variables.

The hereditary terms are calculated using the
prescriptions given in Refs.\,\cite{BS93, BDI95,B96,BIJ02} for tails,
Ref.\,\cite{B98tail} for tails of tails and Ref.\,\cite{B98quad} for tail square.
The complete expression for the energy flux ${\cal F}$ in terms of the
scaled multipoles is given as
\begin{widetext}\begin{eqnarray}
\mathcal{F}=&&\frac{32}{5}\frac{c^5v^{10}}{G}\nu^2\mu_2^2\Bigg\{1+
v^2 \Bigg(-\frac{107}{21}+\frac{55
}{21}\nu+{\hat{\mu}}_3^2 \Bigg[\frac{1367}{1008}-\frac{1367 }{252}\nu\Bigg]+\hat{\epsilon_2}^2\Bigg[\frac{1}{36}-\frac{\nu }{9}\Bigg] \Bigg)
+4\pi v^{3}
+v^{4}\Bigg(\frac{4784}{1323}-\frac{87691}{5292}\nu
\linechange
+\frac{5851}{1323} \nu^2+\hat{\mu}_3^2 \Bigg[-\frac{32807}{3024}+\frac{3515}{72}\nu-\frac{8201}{378} \nu^2\Bigg]+\hat{\mu}_4^2\Bigg[\frac{8965}{3969}-\frac{17930}{1323}\nu+\frac{8965}{441} \nu^2\Bigg]+\hat{\epsilon_2}^2\Bigg[-\frac{17}{504}+\frac{11}{63}\nu-\frac{10}{63} \nu^2\Bigg]
\linechange
+ \hat{\epsilon_3}^2\Bigg[\frac{5}{63}-\frac{10}{21}\nu+\frac{5}{7}\nu^2\Bigg]\Bigg)
+\pi v^{5}\Bigg(-\frac{428}{21}+\frac{178}{21} \nu+\hat{\mu}_3^2 \Bigg[\frac{16403}{2016}-\frac{16403}{504} \nu \Bigg]+\hat{\epsilon_2}^2\Bigg[\frac{1}{18}-\frac{2}{9}\nu\Bigg] \Bigg)
\linechange
+v^6\Bigg(\frac{99210071}{1091475}+\frac{16 \pi^2}{3}
-\frac{1712 }{105}\gamma_E-\frac{856}{105} \log [16 v^2]
+\Bigg[\frac{1650941}{349272}+\frac{41 \pi^2}{48}\Bigg]\nu-\frac{669017}{19404} \nu ^2+\frac{255110}{43659} \nu^3
\linechange
+\hat{\mu}_3^2\Bigg[\frac{7345}{297}-\frac{30103159 }{199584}\nu+\frac{10994153}{49896}\nu^2-\frac{45311}{891}\nu^3\Bigg]+\hat{\mu}_4^2\Bigg[-\frac{1063093}{43659}+\frac{20977942}{130977}\nu-\frac{12978200}{43659}\nu^2
\linechange
+\frac{1568095}{14553}\nu^3\Bigg] +\hat{\mu}_5^2\Bigg[\frac{1002569}{249480}-\frac{1002569}{31185}\nu+\frac{1002569}{12474}\nu^2-\frac{2005138}{31185}\nu^3\Bigg]+\hat{\epsilon}_2^2\Bigg[-\frac{2215}{254016}-\frac{13567}{63504}\nu +\frac{65687}{63504}\nu^2
\linechange
-\frac{853\nu^3}{5292}\Bigg]+\hat{\epsilon_3}^2\Bigg[-\frac{193}{567}+\frac{1304}{567}\nu -\frac{2540}{567}\nu^2+\frac{365}{189}\nu^3\Bigg] +\hat{\epsilon_4}^2\Bigg[\frac{5741}{35280}-\frac{5741 }{4410}\nu+\frac{5741}{1764} \nu^2-\frac{5741}{2205}\nu^3\Bigg]\Bigg)
\linechange
+\pi v^{7}\Bigg(\frac{19136}{1323}-\frac{144449}{2646}\nu+\frac{33389}{2646}\nu^2+\hat{\mu}_3^2\Bigg[-\frac{98417}{1512}+\frac{55457}{192}\nu-\frac{344447}{3024}\nu^2\Bigg]+\hat{\mu}_4^2\Bigg[\frac{23900}{1323}-\frac{47800}{441}\nu
\linechange
+\frac{23900}{147}\nu^2\Bigg]+\hat{\epsilon_2}^2\Bigg[-\frac{17}{252}+\frac{9}{28}\nu-\frac{13}{63}\nu^2\Bigg]+\hat{\epsilon_3}^2\Bigg[\frac{20}{63}-\frac{40}{21}\nu+\frac{20}{7}\nu^2\Bigg]
\Bigg)
\Bigg\},\label{3.5PNEF}
\end{eqnarray}\end{widetext}
where $\hat{\mu}_\ell=\mu_\ell/\mu_2$, $\hat{\epsilon_\ell}=\epsilon_\ell/\mu_2$, Euler constant, $\gamma_E=0.577216$ and $\nu$ is the symmetric
mass ratio defined as the ratio of reduced mass $\mu$ to the total mass
$m$.
As an algebraic check of the result, we recover the GR
results of Ref.~\cite{BIJ02} in the limit $\mu_l\rightarrow 1,
\epsilon_l\rightarrow 1.$
\subsection{Conservative dynamics of the binary}\label{sec2B}
 A model for the conservative dynamics of the binary is also
required to compute the phase evolution of the system. This
enters the phasing formula in two ways. First, the equation of motion of the binary~\cite{BI03CM} in the center-of-mass frame 
is required to compute the derivatives of the multipole moments while calculating the
energy flux.  Second, the expression for the 3PN orbital
energy~\cite{BI03CM, BFIJ02} is necessary to compute the equation of 
energy balance to obtain the phase evolution [see
Eqs.~(\ref{eq:tf})--(\ref{eq:psif}) below]. As the computation of the radiative multipole
moments requires two or more derivative operations, they are implicitly
sensitive to the equation of motion. Hence, formally,  a constraint on the deformation
of the radiative multipole moment does take into account a potential
deviation in the equation of motion from the predictions of GR.

Here however we assume that the conserved energy is the same as in GR.  This
assumption is motivated by practical considerations.
We could have taken a more generic approach by deforming the PN
coefficients in the equation of motion and conserved energy as well.  
As the former is degenerate with the definition of radiative
multipole moments, one would need to consider a parametrized expression
for the conserved energy which will give us a phasing formula with four
additional parameters corresponding to the different PN orders in the
expression for conserved energy. A simultaneous estimation of these
parameters with the multipole coefficients would significantly degrade
the resulting bounds and may not yield meaningful constraints.
However, in the Appendix, we present a parametrized phasing formula
where in addition to the multipole coefficients, various PN-order terms
in the conserved 3PN energy expression are also deformed [see
Eq.~(\ref{eq:psicons}) below]. Interestingly, as can be seen from Eq.~(\ref{eq:psicons}), if there is a modification to the conservative dynamics, they will be fully degenerate with at least one  of the multipole coefficients appearing at the same order.
Due to this degeneracy, such modifications will be detected by this test as 
modifications to ``effective" multipole moments. Further, this degeneracy is not
accidental. It can be shown that by differentiating the expression for the conserved energy, one can derive the energy flux by systematically
accounting for the  equation of motion, including radiation reaction
terms~\cite{IW93,IW95}. We are, therefore, confident that the
power of the proposed test is not diminished by this assumption.
The conserved energy (per unit mass) up to 3PN order is given
by~\cite{DJSdim,DJSequiv,BDE04,ABF01,BI03CM,itoh1,BFIJ02}
%
\begin{eqnarray}\label{eq:Econs}
E(v)&=&-\frac{1}{2} \nu v^2   \left[ 1 - 
\left (\frac{3}{4} + \frac{1}{12} \nu \right ) v^2 - \left( \frac{27}{8} -\frac{19}{8} \nu  + \frac{1}{24}\nu^2 \right) v^4\right.\nonumber\\
  &&\left.- \left\{ \frac{675}{64}
- \left(\frac{34445}{576}-\frac{205 }{96}\pi^2\right)\nu 
+\frac{155 }{96}\nu^2
\right.\right.\nonumber\\
&&\left.\left.+\frac{35 }{5184}\nu^3
\right\} v^6 \right]\,.
\label{eq:energy}
\end{eqnarray}
Using the expressions for the modified
flux and the orbital energy we next proceed to compute the
phase evolution of the compact binary.

\subsection{Computation of the parametrized multipolar phasing formula}
With the parametrized multipolar flux and the energy expressions, we
compute the 3.5PN, nonspinning, frequency-domain phasing formula following
the standard prescription~\cite{DIS00,BIOPS09} by employing the stationary phase
approximation (SPA)~\cite{SatDhu91}.  Consider a GW signal of the form
\begin{equation}
h(t) = \mathcal{A}(t)\cos{\phi(t)}.
\end{equation}
The Fourier transform of the signal will involve an integrand whose
amplitude is slowly varying and whose phase is rapidly oscillating.
In the SPA, the dominant contributions to this integral come from
the vicinity of the stationary points of its phase~\cite{DIS00}. 
As a result the frequency-domain gravitational waveform may
be expressed as
\begin{eqnarray}
&\tilde{h}^{\text{SPA}}(f) &=\frac{\mathcal{A}(t_f)}{\sqrt{\dot {F}(t_f)}}e^{i[\psi_f(t_f)-\pi/4]}\,,\\
&\psi_f(t) &= 2\pi f t- \phi(t),
\end{eqnarray}
where $t_f$ can be obtained by solving $\left . d\psi_f(t)/dt\right |_{t_f} =0$, 
$F(t)$ is the gravitational wave frequency and at $t=t_f$ the GW frequency 
coincides with the Fourier variable $f$. More explicitly,
\begin{eqnarray}
t_f& =&t_{\text{ref}}+ m\int_{v_{f}}^{v_{ref}} \frac{E^\prime
(v)}{\mathcal{F}(v)}dv \,,\label{eq:tf}\\
\psi_f (t_f) &= &2\pi f t_{\text{ref}}-
\phi_{\text{ref}}+2\int_{v_{f}}^{v_{ref}} (v_f^3-v^3)\frac{E^\prime
(v)}{\mathcal{F}(v)}dv\label{spa_taylorf2},\label{eq:psif}
\end{eqnarray}
where $E^\prime(v)$ is the  derivative of the binding energy of the
system expressed in terms of the PN expansion parameter $v$.
Expanding the factor in the integrand in Eq.~(\ref{spa_taylorf2}) as a
PN series and truncating up to 3.5PN order, we obtain the 3.5PN accurate
TaylorF2 phasing formula.

Following the very same procedure, but using Eq.~(\ref{3.5PNEF}) to be the parametrized flux, $\mathcal{F}$, together with the leading quadrupolar order
amplitude (related to the Newtonian GW polarizations), we derive the  standard {\it
restricted} PN waveform in the frequency domain,  which reads as 
\begin{equation}
\tilde{h}(f)=\mathcal{A}\,\mu_2\,f^{-7/6}e^{i\psi(f)}\label{mod-TaylorF2},
\end{equation}
where  $\psi(f)$ is the parametrized multipolar
phasing, $\mathcal{A}=\mathcal{M}_c^{5/6}/\sqrt{30}\pi^{2/3}D_L$; $\mathcal{M}_c=(m_1m_2)^{3/5}/(m_1+m_2)^{1/5}$ 
and $D_L$ are the chirp mass and luminosity distance, respectively, and
$m_1, m_2$ denote the component masses of the binary. Note
the presence of $\mu_2$ in the  GW amplitude, this is due to the mass
quadrupole that contributes to the amplitude at the leading PN order.
If we incorporate the higher-order PN terms in the GW
polarizations~\cite{BIWW96,ABIQ04,BFIS08}, higher-order
multipoles will enter the GW amplitude as well. 

Finally the expression for the 3.5PN frequency-domain phasing $\psi(f)$ is
given by,
\begin{widetext}\begin{eqnarray}
 \psi(f)&=&2\pi f t_c-\frac{\pi}{4}-\phi_c+\frac{3}{128v^5\mu_2^2\nu}\Bigg\{1+
 v^2\Bigg(\frac{1510}{189}-\frac{130}{21}\nu+\hat{\mu}_3^2\Bigg[-\frac{6835}{2268}+\frac{6835}{567}\nu\Bigg]+\hat{\epsilon_2}^2\Bigg[-\frac{5}{81}+ \frac{20}{81}\nu\Bigg]\Bigg)
 \linechange
 -16\pi v^3+
 v^4\Bigg(\frac{242245}{5292}+\frac{4525}{5292} \nu +\frac{145445}{5292} \nu ^2 +{\hat{\mu}_3}^2\Bigg[-\frac{66095}{7056}+\frac{170935 }{3024} \nu-\frac{403405 }{5292}\nu^2\Bigg]+{\hat{\mu}_3}^2\hat{\epsilon_2}^2\Bigg[\frac{6835}{9072}
 \linechange
 -\frac{6835}{1134} \nu +\frac{6835 \nu ^2}{567}\Bigg]
 +\hat{\mu}_3^4\Bigg[\frac{9343445}{508032}-\frac{9343445}{63504} \nu +\frac{9343445}{31752} \nu ^2\Bigg]+\hat{\mu}_4^2\Bigg[-\frac{89650}{3969}+\frac{179300}{1323} \nu  - \frac{89650}{441}\nu^2\Bigg]
 \linechange
 +\hat{\epsilon_2}^2 \Bigg[-\frac{785}{378}+\frac{7115}{756}\nu -\frac{835}{189} \nu ^2\Bigg]
 +\hat{\epsilon_2}^4\Bigg[\frac{5}{648}-\frac{5 }{81} \nu+\frac{10}{81} \nu ^2
 \Bigg]
 +\hat{\epsilon_3}^2\Bigg[-\frac{50}{63}+\frac{100}{21} \nu-\frac{50}{7} \nu ^2\Bigg]
 \Bigg)
 \linechange
 +\pi v^5 \Big(3 \log\left[\frac{v}{v_{\rm LSO}}\right]+1\Big) \Bigg(\frac{80}{189} \Big[151-138 \nu \Big]
 -\frac{9115}{756}\hat{\mu}_3^2\Big[ 1-4 \nu\Big]- \frac{20}{27} \hat{\epsilon_2}^2\Big[ 1-4 \nu\Big]
 \Bigg)+v^6\Bigg(\frac{5334452639}{2037420}
 \linechange
 -\frac{640}{3}\pi^2-\frac{6848}{21}\gamma_E -\frac{6848}{21} \log[4 v]-\Bigg[\frac{7153041685}{1222452}-\frac{2255}{12}\pi^2\Bigg]\nu +\frac{123839990}{305613}\nu^2+ \frac{18300845}{1222452}\nu^3
 \linechange
 +\hat{\mu}_3^2\Bigg[
 -\frac{4809714655}{29338848}+\frac{8024601785}{9779616} \nu-\frac{19149203695}{29338848} \nu^2-\frac{190583245}{7334712} \nu^3\Bigg]
 +\hat{\mu}_3^2\hat{\epsilon_2}^2 \Bigg[-\frac{656195}{95256}+\frac{229475\nu}{3888}
 \linechange
 -\frac{3369935\nu ^2}{23814}+\frac{82795 \nu ^3}{1323}
 \Bigg]
 +\hat{\mu}_3^2\hat{\epsilon_2}^4\Bigg[\frac{6835}{108864}-\frac{6835
 }{9072} \nu+\frac{6835}{2268} \nu^2-\frac{6835}{1701} \nu^3
 \Bigg]
 +\hat{\mu}_3^2\hat{\epsilon_3}^2\Bigg[-\frac{34175}{7938
 }+\frac{170875}{3969} \nu 
 \linechange
 -\frac{375925}{2646}\nu ^2+\frac{68350}{441} \nu ^3
 \Bigg]
 +\hat{\mu}_3^2\hat{\mu}_4^2\Bigg[-\frac{61275775}{500094}+\frac{306378875 
 }{250047}\nu-\frac{674033525}{166698} \nu^2 +\frac{122551550}{27783} \nu^3
 \Bigg]
 \linechange
 +\hat{\mu}_3^4\Bigg[\frac{868749005}{10668672}-\frac{2313421945}{3556224} \nu+\frac{191974645}{148176}\nu^2+\frac{9726205}{666792} \nu^3\Bigg]
 +\hat{\mu}_3^4\hat{\epsilon_2}^2\Bigg[\frac{9343445}{3048192} -\frac{9343445}{254016}\nu
 \linechange
 +\frac{9343445}{63504} \nu^2-\frac{9343445}{47628} \nu^3
 \Bigg]
 + \hat{\mu}_3^6\Bigg[\frac{12772489315}{256048128}-\frac{12772489315 }{21337344} \nu+\frac{12772489315}{5334336} \nu^2-\frac{12772489315}{4000752} \nu^3\Bigg]
 \linechange
 +\hat{\mu}_4^2\Bigg[-\frac{86554310}{916839}+\frac{553387330}{916839} \nu-\frac{289401650}{305613} \nu^2-\frac{4322750}{101871} \nu ^3\Bigg]+ \hat{\mu}_4^2\hat{\epsilon_2}^2\Bigg[-\frac{89650}{35721}+\frac{896500}{35721}\nu
 \linechange
 -\frac{986150}{11907}\nu^2+\frac{358600 \nu ^3}{3969}
 \Bigg]
 +\hat{\mu}_5^2\Bigg[ \frac{1002569}{12474}-\frac{4010276}{6237} \nu+\frac{10025690}{6237} \nu^2-\frac{8020552 }{6237}\nu^3\Bigg]
 \linechange
 +\hat{\epsilon_2}^2\Bigg[\frac{3638245}{190512}-\frac{2842015}{31752}\nu+\frac{760985}{13608}\nu^2-\frac{328675}{23814} \nu^3\Bigg] +\hat{\epsilon_2}^2\hat{\epsilon_3}^2\Bigg[-\frac{50}{567}+\frac{500}{567} \nu -\frac{550}{189} \nu^2+\frac{200}{63} \nu ^3\Bigg ]
 \linechange
 +\hat{\epsilon_2}^4\Bigg[-\frac{265}{1512}+\frac{20165}{13608} \nu-\frac{5855}{1701}\nu^2+\frac{310}{243} \nu ^3\Bigg]
 +\hat{\epsilon_2}^6\Bigg[\frac{5}{11664}-\frac{5}{972} \nu +\frac{5}{243} \nu ^2 -\frac{20}{729} \nu^3\Bigg]
 \linechange
 +\hat{\epsilon_3}^2\Bigg[\frac{27730}{3969}-\frac{179990}{3969} \nu+\frac{341450}{3969} \nu^2- \frac{51050}{1323}\nu^3\Bigg]
 +\hat{\epsilon_4}^2\Bigg[\frac{5741}{1764}-\frac{11482}{441} \nu+\frac{28705 }{441}\nu ^2 -\frac{22964}{441}\nu ^3\Bigg]\Bigg)
 \linechange
 +\pi v^7\Bigg(\frac{484490}{1323}-\frac{141520}{1323}\nu+\frac{442720}{1323}\nu^2
 +\hat{\mu}_3^2\Bigg[-\frac{88205}{2352}+\frac{63865}{252}\nu-\frac{182440 }{441}\nu^2\Bigg]
 +\hat{\mu}_3^2\hat{\epsilon_2}^2\Bigg[\frac{54685}{9072}
 \linechange
 -\frac{54685}{1134}  \nu+\frac{54685 }{567}  \nu^2\Bigg]
 +\hat{\mu}_3^4\Bigg[\frac{6835}{254016}-\frac{6835}{31752}  \nu+\frac{6835}{15876}\nu^2\Bigg]
 +\hat{\mu}_4^2\Bigg[-\frac{400}{3969}+\frac{800}{1323} \nu -\frac{400}{441}\nu ^2\Bigg]
 \linechange
 +\hat{\epsilon_2}^2\Bigg[-\frac{1570}{63}+\frac{7220}{63}\nu -\frac{3760}{63}\nu ^2\Bigg]
 +\hat{\epsilon_3}^2\Bigg[-\frac{400 }{63}+\frac{800}{21} \nu -\frac{400}{7}\nu ^2
 \Bigg]
 +\hat{\epsilon_2}^4\Bigg[\frac{10}{81}-\frac{80}{81} \nu +\frac{160 }{81}\nu ^2\Bigg]
 \Bigg)\Bigg\}\,.\label{3.5PNphasing}
\end{eqnarray}
\end{widetext}
This parametrized multipolar phasing formula constitutes one of the most important 
results of the paper and forms the basis for the analysis which follows.

\subsection{Multipole structure of the post-Newtonian phasing formula}

We summarize in Table~\ref{Non-GR_Params_at_every_PN_order} the
multipole structure of the PN phasing formula based on Eq.~(\ref{3.5PNphasing}).
The various multipoles which contribute to the different PN phasing
terms are listed. The main features are as follows. As we go
to higher PN orders, in addition to the higher-order multipoles making an
appearance, higher-order PN corrections to the lower-order multipoles also
contribute. For example, the mass quadrupole and its corrections (terms
proportional to $\mu_2$)
appear at every PN order starting from 0PN. The 1.5PN and 3PN log terms
contain only $\mu_2$ and are due to the leading-order tail
effect~\cite{BS93} and tails-of-tails effect~\cite{B98tail},
respectively. The 3PN nonlogarithmic term contains all seven 
multipole coefficients.

Due to the aforementioned structure, it is evident that if one of the multipole moments is different from
GR, it is likely to affect the phasing coefficients at more than one PN
order. For instance, a deviation in $\mu_2$ could result in a dephasing of
each of the PN phasing coefficients. There are seven independent
multipole coefficients which determine eight PN coefficients. The eight
equations which relate the phasing terms to the multipoles are
inadequate to extract all seven multipoles.
This is because three of the eight equations relate the PN
coefficients only to $\mu_2$, and another two relate the 1PN and 2.5PN
logarithmic terms to a set of three multipole coefficients $\{\mu_2,
\mu_3, \epsilon_2\}$. It turns out that, in principle,  by independently measuring 
the eight PN coefficients, we can measure all the multipoles except
$\mu_5$ and $\epsilon_4$. It is well known that measuring all eight
phasing coefficients together provides very bad bounds~\cite{AIQS06a,AIQS06b}.
The version of the parametrized tests of post-Newtonian theory, where we
vary only one parameter at a time~\cite{AIQS06b,TIGER}, cannot be mapped
to the multipole coefficients, as varying multipole moments will cause
more than one PN order to change, which conflicts with the original
assumption. 

Though mapping the space of PN coefficients to that of the multipole
coefficients is not possible,  it is possible to relate the multipole
deformations to that of the parametrized test.  If, for instance, 
$\mu_2$ is different from GR, it can lead to dephasing in one or more 
of the PN phasing terms depending on what the correction is to the mass
quadrupole at different PN orders. Based on the
multipolar structure, This motivates us to perform
parametrized tests of PN theory while varying simultaneously certain PN coefficients\footnote{We thank Archisman Ghosh for pointing out this possibility 
to us.}. 
\begin{center}
	\begin{table}
		\begin{tabular}{||c | c | c ||} 
			\hline
			PN order  & frequency dependences  & Multipole coefficients  \\ [0.5ex] 
			\hline\hline
			0 PN & $f^{-5/3}$ & $\mu_2$  \\ 
			\hline
			1 PN & $f^{-1}$  &  $\mu_2$, $\mu_3$, $\epsilon_2$ \\
			\hline
			1.5 PN & $f^{-2/3}$ &  $\mu_2$ \\
			\hline
			2 PN & $f^{-1/3}$ &  $\mu_2$, $\mu_3$, $\mu_4$, $\epsilon_2$, $\epsilon_3$  \\
			\hline
			2.5 PN log & $\log f$  & $\mu_2$, $\mu_3$, $\epsilon_2$ \\
			\hline
			3 PN & $f^{1/3}$ & $\mu_2$, $\mu_3$, $\mu_4$, $\mu_5$, $\epsilon_2$, $\epsilon_3$, $\epsilon_4$  \\
			\hline
			3 PN log &$f^{1/3}\log f$& $\mu_2$  \\
			\hline
			3.5 PN & $f^{2/3}$ &  $\mu_2$, $\mu_3$, $\mu_4$, $\epsilon_2$, $\epsilon_3$\\ [1ex] 
			\hline
		\end{tabular}
		\caption{Summary of the multipolar structure of the PN
phasing formula. The contributions of various multipoles to different phasing
coefficients and their frequency dependences are tabulated. Following the definitions introduced in the
paper, $\mu_l$ are associated to the deformations of 
mass-type multipole moments and $\epsilon_l$ refer to the deformations of current-type
multipole moments.}
		\label{Non-GR_Params_at_every_PN_order}
	\end{table}
\end{center}

\section{Parameter estimation of the multipole coefficients}\label{Param_est}
In this section, we will set up the parameter estimation problem to measure the 
multipolar coefficients and present our forecasts for Advanced LIGO, the Einstein Telescope, Cosmic 
Explorer and LISA.  Using the frequency-domain gravitational waveform, we study
how well the current and future generations of GW detectors can probe the
multipolar structure of GR. To quantify this, we derive the projected accuracies
with which various multipole moments may be measured for various
detector configurations by using standard parameter estimation
techniques. Following the philosophy of Refs.\,\cite{AIQS06a,MAIS10,TIGER}, while
computing the errors we consider the deviation of only one multipole at a time. 

An ideal test would have been where all the coefficients are varied at the
same time, but this would lead to almost no meaningful constraints
because of the strong degeneracies among different coefficients. 
The proposed test, however, would not affect our ability to detect a 
potential deviation because in the multipole structure, a deviation of 
more than one multipole coefficient would invariably show up in the set of tests
performed by varying one coefficient at a time~\cite{MAIS10,TIGER,Li:2011cg,Meidam:2017dgf}.

We first use the Fisher information matrix approach to derive the errors on the 
multipole coefficients. The Fisher matrix is a useful semianalytic method which uses a
quadratic fit to the log-likelihood function to derive the $1\sigma$ error bars on 
the parameters of the signal~\cite{Rao45,Cramer46,CF94,AISS05}. 
Given a GW signal $\tilde{h}(f;\vec{\theta})$, which is described by the
set of parameters ${\vec \theta}$, the Fisher information matrix is defined as
\begin{equation}
\label{eq:fisher}
\Gamma_{mn}=\langle\tilde{h}_m,\tilde{h}_n\rangle,
\end{equation}
where $\tilde{h}_m=\partial\tilde{h}(f;\vec{\theta})/\partial\theta_m$, and the angular bracket, $\langle...,...\rangle$,  denotes the noise-weighted inner product 
defined by 
\begin{equation}
\label{eq:innerproduct}
\langle a,b\rangle=2\int_{f_{\rm low}}^{\rm f_{\rm high}}\frac{a(f)\,b^*(f)+a^*(f)\, b(f)}{S_h(f)}\,df \,.
\end{equation}
Here $S_h(f)$ is the one-sided noise power spectral density (PSD) of the detector and $[f_{\rm low}, f_{\rm high}]$ are the lower and upper limits of integration.
The variance-covariance matrix is defined by the inverse of the Fisher 
matrix, $$C^{mn}=(\Gamma^{-1})^{mn},$$ where the diagonal components, $C^{mm}$, are 
the variances of $\theta^m.$ The $1\sigma$ errors on $\theta^m$ are, therefore, given as
\begin{equation}
\sigma^m = \sqrt{C^{mm}} \,.
\end{equation}

Since Fisher-matrix-based estimates are only reliable in the high signal-to-noise ratio
limit~\cite{CF94,BalSatDhu95,Vallisneri07}, we spot check representative cases for
consistency, with estimates based on a Bayesian inference algorithm that uses an 
MCMC method to sample the likelihood function. This method is not limited by the 
quadratic approximation to the log-likelihood and hence is considered to be a more reliable estimate of measurement accuracies one might have in a real experiment. In this 
method we compute the probability distribution for the parameters implied by a signal 
$h(t)$ buried in the Gaussian noise $d(t)=h(t)+n(t)$ while incorporating our prior 
assumptions about the probability distribution for the parameters. Bayes’ rule states that the probability distribution for a set of model parameters 
$\vec{\theta}$ implied by data $d$ is
\begin{equation}
p(\vec{\theta}|d) = \frac{p(d|\vec{\theta})\, p(\vec{\theta})}{p(d)}\,,
\end{equation} 
where $p(d|\vec{\theta})$ is called the {\it likelihood} function, which gives the 
probability of observing data $d$ given the model parameter $\vec{\theta}$, defined as
\begin{equation}
p(d|\vec{\theta}) = \exp \Bigg[-\frac{1}{2} \, \int^{f_{\rm high}}_{f_{\rm low}} \frac{|\tilde{d}(f)-\tilde{h}(f;\vec{\theta})|^2}{S_h(f)} \, df\Bigg]\,,
\label{eq_likelihood}
\end{equation} 
where $\tilde{d}(f)$ and $\tilde{h}(f; \vec{\theta})$ are the Fourier transforms of $d(t)$ and $h(t)$, respectively. 
$p(\vec{\theta})$ is the {\it prior probability distribution} of parameters $\vec{\theta}$ and
$p(d)$ is an overall normalization constant known as the {\it evidence},
\begin{equation}
p(d) = \int p(d|\vec{\theta})\, p(\vec{\theta}) \, d\vec{\theta}\,.
\end{equation}   
In this paper, we use uniform prior on all the parameters we are
interested in and used {\tt python}-based MCMC sampler {\tt emcee}
~\cite{emcee} to sample the likelihood surface and  get the  posterior distribution for all the parameters.

We use the noise PSDs of advanced LIGO (aLIGO) and Cosmic Explorer-wide band (CE-wb)
\cite{CE_WB}, Einstein Telescope-D (ET-D)~\cite{ET-D} as representatives of the current and next generations of
ground-based GW interferometers and LISA. We use the noise PSD given in Ref.~\cite{ET-D} for ET-D, analytical fits of PSDs given
in Refs.~\cite{Ajith2011b} and \cite{Babak2017} for aLIGO and LISA respectively, 
and the following fit for the CE-wb noise PSD, 
\begin{eqnarray}
 S_h(f) &=& 5.62\times10^{-51} + 6.69\times10^{-50}f^{-0.125} + \frac{7.80\times10^{-31}}{f^{20}} \nonumber \\
          &+& \frac{4.35\times10^{-43}}{f^6} + 1.63\times10^{-53}\,f + 2.44\times10^{-56}\,f^2 \nonumber \\
          &+& 5.45\times10^{-66}\,f^5\,\, {\rm Hz^{-1}}\,,
\end{eqnarray}
where $f$ is in units of Hz.  We compute the Fisher matrix (or likelihood in
the Bayesian framework) considering the signal to be described by the set of
parameters $\{{\rm ln}\mathcal{A}, {\rm ln}M_c, {\rm ln}\nu, t_c, \phi_c\}$ and
the additional parameter $\mu_l$ or $\epsilon_l$. In order to compute the inner
product using Eq.~(\ref{eq:innerproduct}), we assume $f_{\rm low}$ to be $20$, $1$, $5$ and $10^{-4}$ Hz for the aLIGO, ET-D, CE-wb and LISA noise PSDs respectively.
We choose $f_{\rm high}$ to be the frequency at the last stable circular orbit
of a Schwarzschild BH with a total mass $m$ given by $f_{\rm LSO}=1/(\pi\,m\,
6^{3/2})$ for the aLIGO, ET-D and CE-wb noise PSDs.  For LISA, we choose the upper cut off
frequency to be the minimum of $[0.1, f_{\rm LSO}]$. Additionally, LISA being a triangular shaped detector we multiply our gravitational waveform by a factor of $\sqrt{3}/2$ while calculating the Fisher matrix for LISA.

\begin{figure*}[t]
	\centering
	\includegraphics[scale=0.42]{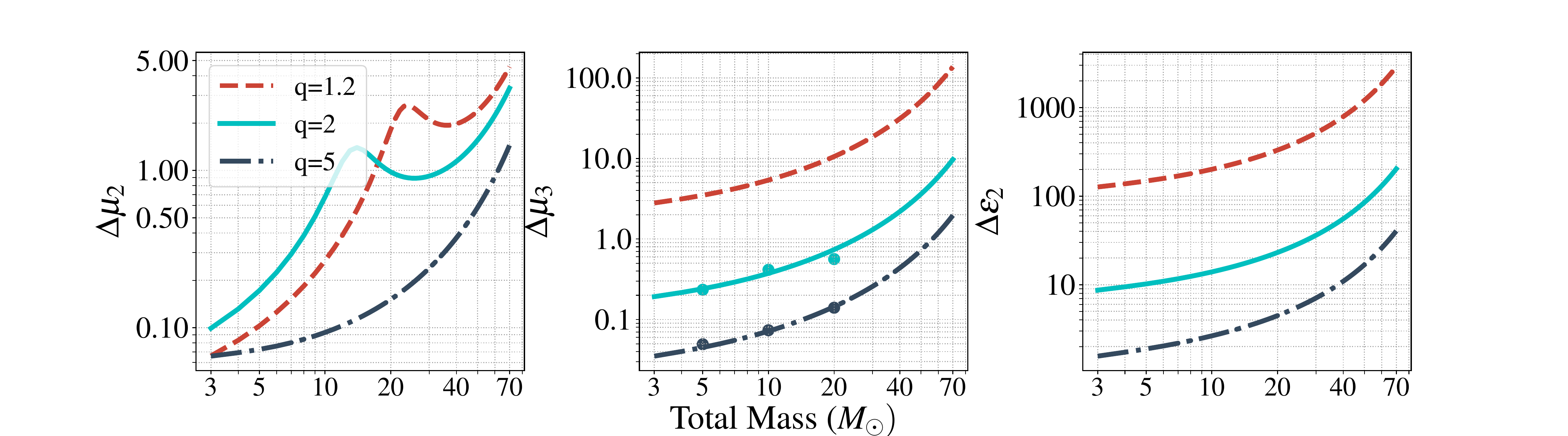}
	\caption{Projected $1\sigma$ errors on $\mu_2$, $\mu_3$ and $\epsilon_2$ as a functions of the total mass for the aLIGO noise PSD. Results from Bayesian analysis using MCMC sampling are given as dots showing good agreement. All the sources are considered to be at a fixed luminosity distance of 100 Mpc.}\label{fig1}
\end{figure*}
\begin{figure*}[t]
	\includegraphics[scale=1]{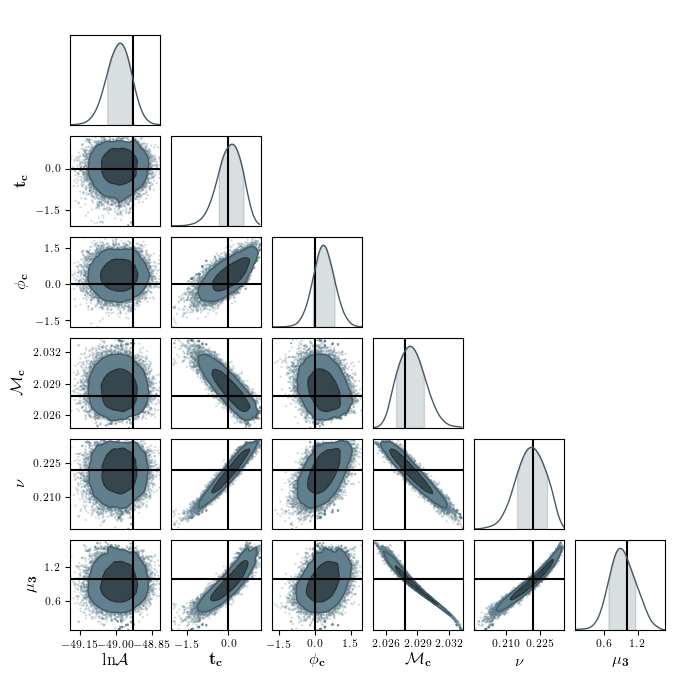}
	\caption{The posterior distributions of all six parameters $\{{\rm ln}\mathcal{A}, t_c, \phi_c,\mathcal{M}_c, \nu,\mu_3\}$ and their corresponding contour plots obtained from the MCMC experiments (see Sec.~\ref{Param_est} for details) for a compact binary system at a distance of 100 Mpc with $q=2$, $m=5 \text{ M}_\odot$ using the noise PSD of aLIGO. The darker shaded regions in the posterior distributions as well as in the contour plots shows the $1\sigma$ bounds on the respective parameters.\label{figMCMC_aLIGO}}
\end{figure*}

\begin{figure*}[t]
	\includegraphics[scale=0.42]{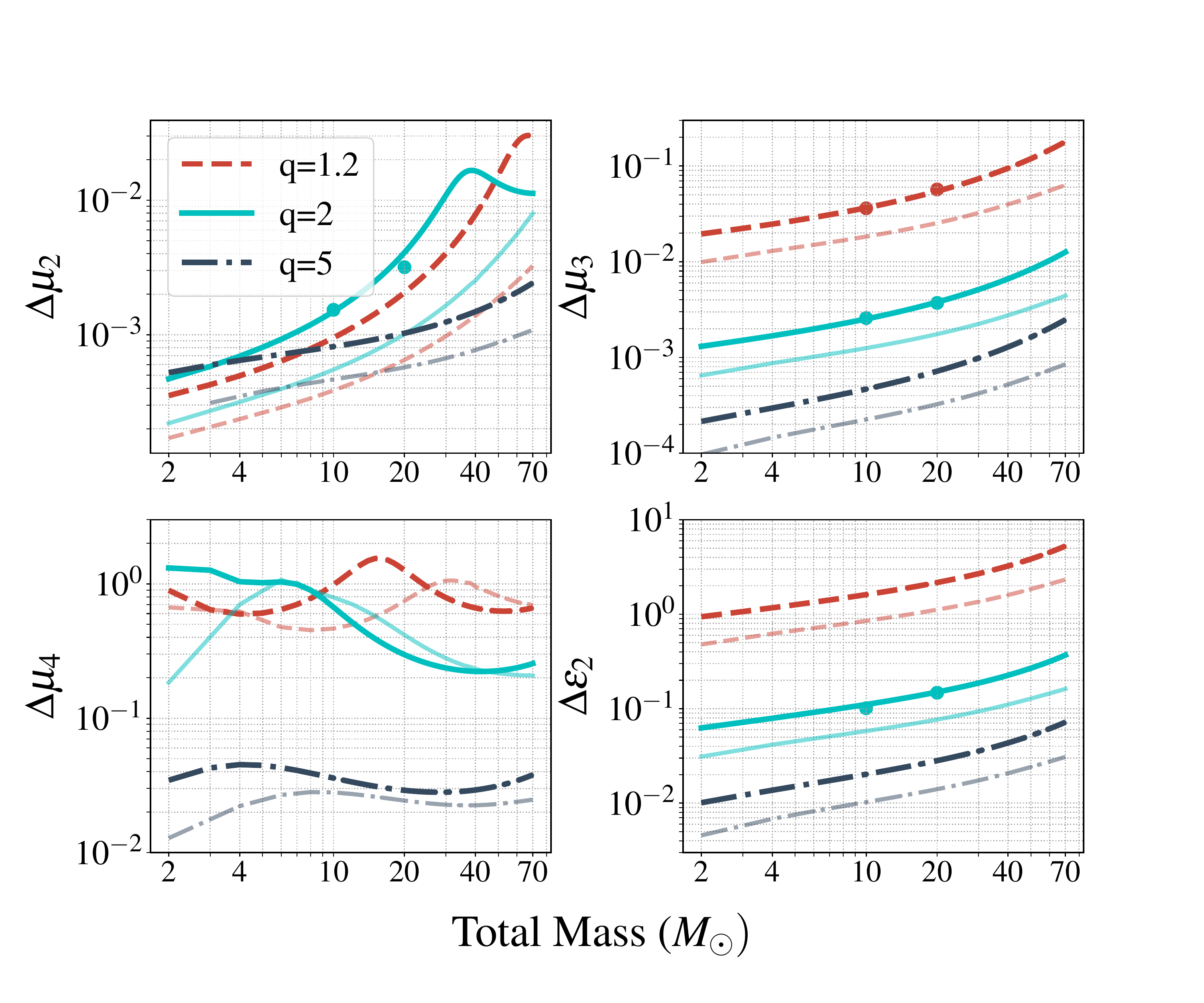}
	\caption{Dark shaded curves correspond to the projected $1\sigma$ error bars on $\mu_2$, $\mu_3$, $\mu_4$ and $\epsilon_2$ using the proposed CE-wb noise PSD as a function of the total mass, where as lighter shades denote the bounds obtained using the ET-D noise PSD. All the sources are considered to be at a fixed luminosity distance of 100 Mpc. The higher-order multipole moments such as $\mu_4$ and $\epsilon_2$ cannot be measured well using aLIGO and hence it may be a unique science goal of the third-generation detectors.\label{fig2}}
\end{figure*}
\begin{figure*}[t]
	\includegraphics[scale=.9]{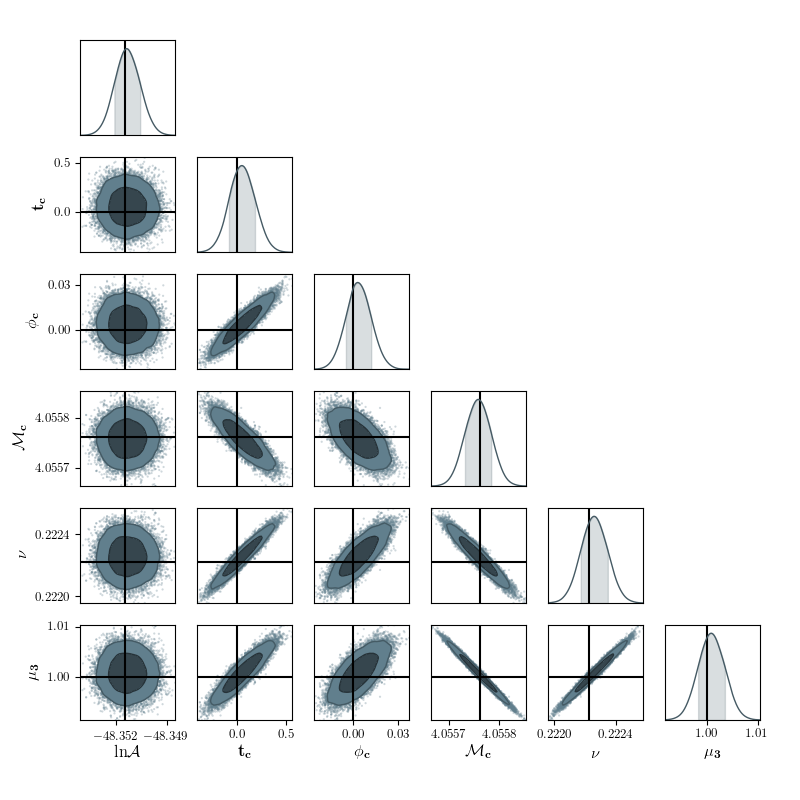}
	\caption{The posterior distributions of all six parameters $\{{\rm ln}\mathcal{A}, t_c, \phi_c,\mathcal{M}_c, \nu,\mu_3\}$ and their corresponding contour plots obtained from the MCMC experiments (see Sec.~\ref{Param_est} for details) for a compact binary system at a distance of 100 Mpc with $q=2$, $m=10 \text{ M}_\odot$ using the noise PSD of CE-wb. The darker shaded region in the posterior distributions as well as in the contour plots shows the $1\sigma$ bounds on the respective parameters.\label{figMCMC_CE}}
\end{figure*}

All of the parameter estimations for aLIGO, CE-wb and LISA, that we carry out here, assume
detections of the signals with a single detector, whereas for ET-D, due to its triangular shape, we consider the noise PSD to be enhanced roughly by a factor of 1.5.  As our aim is to
estimate the intrinsic parameters of the signal, which directly affect
the binary dynamics, the single detector estimates are good enough for our
purposes and a network of detectors may improve it by the square root of
the number of detectors. Hence the reported errors are likely to give rough,
but conservative, estimates of the expected accuracies with which the multipole
coefficients may be estimated.

\section{ Results and Discussion}\label{results}
In this section, we report the $1\sigma$ measurement errors on the multipole 
coefficients introduced in the previous section, obtained using the
Fisher matrix as well as Bayesian analysis and discuss their implications.

Our results for the four different detector configurations are presented in
Figs.~\ref{fig1}, ~\ref{fig2} and \ref{fig4} which show the errors on
the various multipole coefficients ${\mu_l}, {\epsilon_l}$ for aLIGO,  ET-D, CE-wb
and LISA, respectively.  For all of these estimates we consider the sources
at fixed distances. In addition to the intrinsic parameters there are four more (angular) parameters 
that are needed to completely specify the gravitational waveform. More specifically one needs two angles to 
define the location of the source on the sky and another two angles to specify the orientation 
of the orbital plane with respect to the detector
plane~\cite{SathyaSchutzLivRev09}.
Since we are using a pattern-averaged waveform~\cite{DIS00} (i.e., a
waveform averaged  over all four angles), the luminosity
distance can be thought of as an {\it effective} distance  which we assume 
to be 100 Mpc for aLIGO, ET-D and CE-wb, and 3 Gpc for LISA. For aLIGO, ET-D and CE-wb, 
we explore the bounds for the binaries with a total mass in the range 
[1,70] $\text{M}_\odot$ and for LISA detections in the range $[10^5,10^7]$ $\text{M}_\odot$.

\subsection{Advanced LIGO}

In Fig.~\ref{fig1} we show the projected $1$-$\sigma$ errors on the three
leading-order multipole moments, $\mu_2, \mu_3$ and $\epsilon_2,$ as a function
of the total mass of the binary for the aLIGO noise PSD using the Fisher matrix.  Different curves are for different mass
ratios, $q=m_1/m_2=1.2$ (red), $2$ (cyan) and $5$ (blue).  For the multipole 
coefficients considered, low-mass systems obtain the smallest errors and hence the
tightest constraints. This is expected as low-mass systems live longer in the
detector band and have larger number of cycles, thereby allowing us to measure
the parameters very well.  The bounds on $\mu_3$ and $\epsilon_2$, associated
with the mass octupole and current quadrupole, increase monotonically with the total
mass of the system for a given mass ratio.  However, the bounds on $\mu_2$ 
show a local minimum in the intermediate-mass regime for smaller mass ratios. 
This is because, unlike  other multipole parameters, $\mu_2$ appears both in the 
amplitude and the phase of the signal. The derivative of the
waveform with respect to $\mu_2$ has contributions from both the amplitude
and phase. Schematically, the Fisher matrix element is given by
\begin{equation}
\Gamma_{\mu_2\mu_2}\sim\int_{f_{\rm low}}^{f_{\rm high}}\frac{{\cal A}^2
f^{-7/3}}{S_h(f)}\left(1+\mu_2^2\psi'^2\right)\,df,
\end{equation}
where $\psi'=\partial \psi/\partial \mu_2$. As the inverse of this term
dominantly determines the error on $\mu_2$, the local minimum is a result of the
trade-off between the contributions from the amplitude and the phase of the
waveform. Interestingly, as we go to higher mass ratios, this feature disappears
resulting in a monotonically increasing curve (such as for $q=5$).

We find that the mass multipole moments $\mu_2$ and $ \mu_3$ are much better
estimated as compared to the current multipole moment $\epsilon_2$.
Another important feature is that the bounds $\mu_3$ and $\epsilon_2$
are  worse for equal-mass binaries. The mass-octupole and current-quadrupole
are odd-parity multipole moments (unlike, say,  the mass quadrupole which is
even)\footnote{Mass-type multipoles with even $l$ and current-type
moments with odd $l$ are considered `even' and odd $l$ mass multipoles and
even $l$ current moments are `odd'.}. Every odd-parity multipole moment
comes with a mass asymmetry factor $\sqrt{1-4\nu}$ that vanishes
in the equal-mass limit, and hence the errors diverge. Consequently,  
the Fisher matrix  becomes badly conditioned and the precision with which we recover
these parameters appears to become very poor, but this is an artifact of the Fisher matrix. 

In order to cross-check the validity of the Fisher-matrix-based
estimates, we performed a Bayesian analysis to find the posterior distribution of
the three multipole parameters, for the same systems as in the Fisher matrix analysis. 
Moreover we considered a flat prior probability distribution for all six parameters 
$\{{\rm ln}\mathcal{A}, M_c, \nu, t_c,\phi_c, \mu_\ell \text{ or } \epsilon_\ell\}$ in a large enough range around their respective injection values.  
Given the large number of iterations, once the MCMC chains are stabilized, we find good agreements with the Fisher estimates as in the case of $\mu_3$ for $q=2$ and $5$, shown in Fig.~\ref{fig1}. 
As an example, we present our results from the MCMC analysis for $\mu_3$
with $m=5 \text{ M}_\odot$ and $q=2$, in the corner plots
in Fig.~\ref{figMCMC_aLIGO}. In Fig.~\ref{fig1} we see that the
$1\sigma$ errors in $\mu_3$ from the Fisher analysis agree very well with
the MCMC results for $q=2$ and $5.$ We did not find such an agreement 
for $q=1.2$. We suspect that this is because for comparable-mass systems the likelihood function, defined in Eq.~(\ref{eq_likelihood}), becomes shallow and 
it is computationally very difficult to find its maximum given a finite number 
of iterations. As a result, the MCMC chains did not converge  and $1\sigma$
bounds cannot be trusted for such cases. We find the nonconvergence of MCMC chains for all of the
cases of $\mu_2$ and $\epsilon_2$ and hence we do not show those results
in Fig.~\ref{fig1}. To summarize, our findings indicate that one can only measure $\mu_2$ and $\mu_3$ with a good enough accuracy using aLIGO detectors.

\subsection{Third-generation detectors}

Third-generation detectors such as CE-wb (and ET-D) can place much better bounds on $\mu_2,
\mu_3$ and $\epsilon_2$ compared to aLIGO. Additionally, they can also  
measure $\mu_4$ with reasonable accuracy,  as shown by the darker (and lighter) shaded curves in Fig.~\ref{fig2}.
The bounds on $\mu_2$, $\mu_3$ and $\epsilon_2$ show similar trends as in
the case of aLIGO except the accuracy of the parameter estimation is  much
better overall.   
For a few cases in low-mass regime, $\mu_2$ and $\mu_4$ are better estimated for comparable-mass binaries (i.e., $q=1.2$). We also find that the bounds (represented by the lighter shaded curves in Fig.~\ref{fig2}) obtained by using the ET-D noise PSD are even better than the bounds from CE-wb, though the other features are more or less similar for both of the detectors. This improvement in the precision of measurements is due to two reasons. The triangular shape of ET-D enhances the sensitivity roughly by a factor of 1.5 and its sensitivity is much better than CE-wb in the low-frequency region.

For a few representative cases, we compute the errors in $\mu_2$, $\epsilon_2$ and $\mu_3$ using Bayesian analysis and the results are shown as
dots with the same color in Fig.~\ref{fig2}. The MCMC results are in good
agreement with the Fisher matrix results. Unlike the aLIGO PSD, for CE-wb the MCMC chains converge quickly 
in the case of $\mu_2$ and $\epsilon_2$ because of the high
signal-to-noise-ratios, which naturally lead to high likelihood values.  As a
result, it becomes relatively easier for the sampler to find the global
maximum of the likelihood function in relatively fewer iterations. We also show an example corner plot for the CE-wb PSD with $q=2$, $m=10 \text{ M}_\odot$ in Fig.~\ref{figMCMC_CE}.

\begin{figure*}[t]\centering
	\includegraphics[scale=0.32]{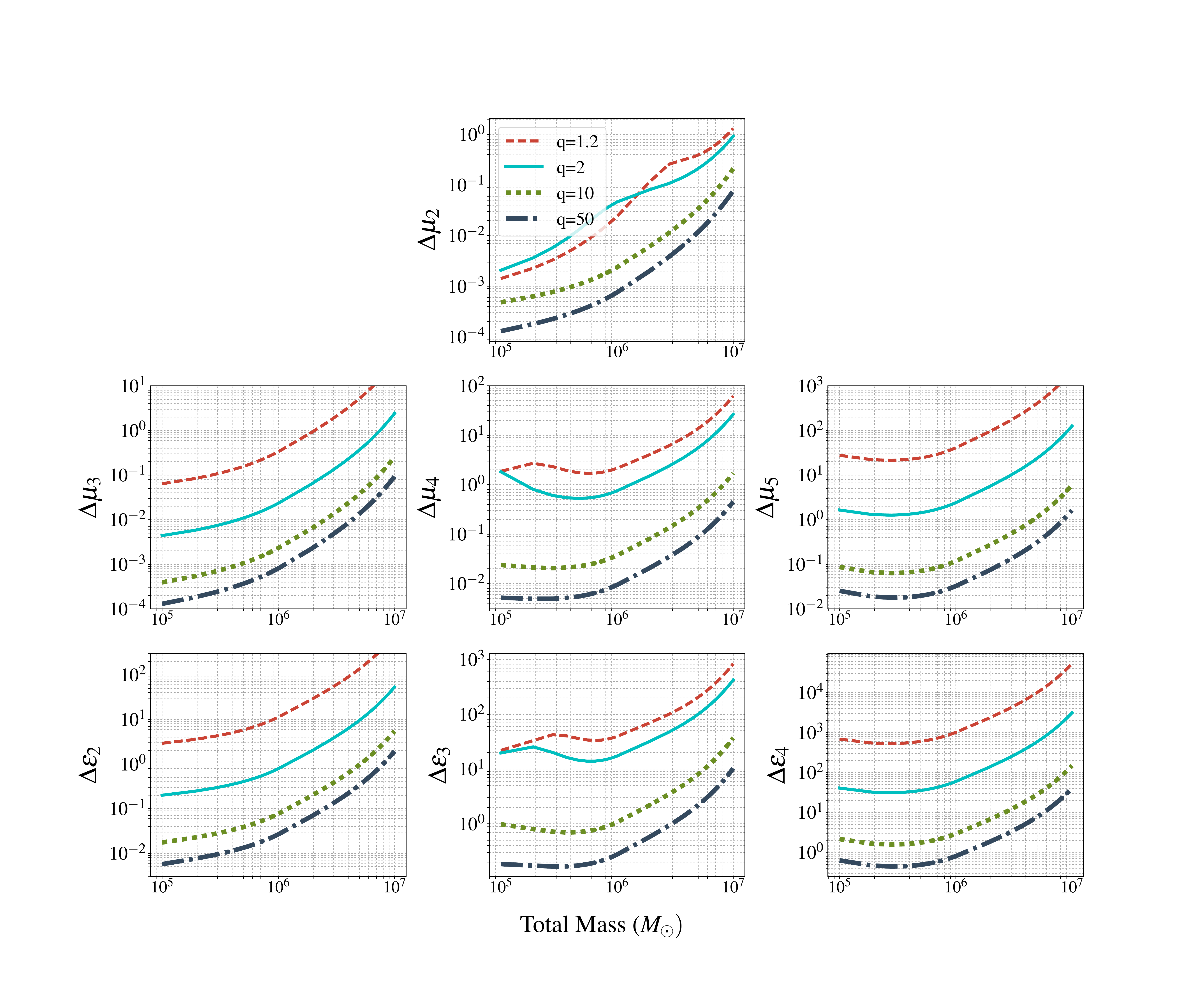}
	\caption{Projected constraints on various multipole coefficients
		using LISA sensitivity, as a function of the total mass of the binary. All the sources are considered to be at a fixed luminosity distance of 3 Gpc. LISA can measure all seven multipoles which contribute to the phasing and hence will
		be able to place extremely stringent bounds on the multipoles of the
		compact binary gravitational field.\label{fig4}}
\end{figure*}

\subsection{Laser Interferometer Space Antenna}
In this section, we discuss the projected errors on various multipole coefficients  for the LISA detector.  Here we consider four different mass ratios, $q=1.2$ (red), $2$ (cyan), $10$ (blue) and $50$ (green).  The first three are representatives of
comparable-mass systems, while $q=50$ refers to the intermediate-mass
ratio systems. We do not consider here the extreme-mass-ratio-systems, the analysis of these systems needs phasing information at much higher
PN orders such as in Ref.~\cite{Fujita12} which is beyond
the scope of the present work. Moreover, in such systems, the motion of
the smaller BH around the central compact object is expected to help us
understand the multipolar structure of the central object and test its
BH nature~\cite{Ryan97}. This is quite different from our
objective here which is to use GW observations to understand the
multipole structure of the gravitational field of the two-body problem
in GR.   The $q=50$ case, in fact, falls in between these two
classes and hence 
has a cleaner interpretation in our framework.

In Fig.~\ref{fig4} we show the projected bounds from the observations of
supermassive BH mergers detectable by the space-based LISA observatory. 
The error estimates for multipole moments with LISA are similar to that
of CE-wb for mass ratios $q=1.2$, $2$.  For $q=10$
all the parameters except $\epsilon_4$ are estimated very well. 
For $q=50$, we find that LISA will be able to measure all seven multipole
coefficients with good accuracy. It is not entirely clear whether
PN model is accurate enough for the detection and parameter estimation of 
supermassive binary BHs with $q=50$, for which the number of GW
cycles could be an order of magnitude higher than it is for equal-mass
configurations.  However our findings carry important as they point to the huge
potential such systems have for fundamental physics.

To summarize, we find, in general,  that even-parity multipoles (i.e., $\mu_2$
and $\mu_4$) are better measured when the binary constituents are of equal or
comparable masses, whereas the odd multipoles (i.e., $\mu_3$, $\mu_5$,
$\epsilon_2$ and $\epsilon_3$) are better measured when the binary has mass
asymmetry. This is because the even multipoles are proportional to the symmetric
mass ratio $\nu,$ whereas the odd ones are proportional to the mass asymmetry
$\sqrt{1-4\nu},$ which identically vanishes for equal-mass systems [see, e.g.,
Eq. (4.4) of Ref.~\cite{BDI95}].

\section{Summary and Future directions}\label{sec:conclusion}

We have proposed a novel way to test for possible deviations from GR using GW
observations from compact binaries by probing the multipolar structure
of the GW phasing in any alternative theories of gravity. We compute a
parametrized multipolar GW phasing formula that can be used to probe
potential deviations from the multipolar structure of GR.
Using the Fisher information matrix and Bayesian parameter estimation, we
predict the accuracies with which the multipole coefficients could be
measured from GW observations with present and future detectors. 
We find that the space mission LISA, currently under development,
can measure all the multipoles of the compact binary system. Hence this will be among the
unique fundamental science goals LISA can achieve.

In deriving the parametrized multipolar phasing formula, we have
assumed that the conservative dynamics of the binary follow the
predictions of GR. In the Appendix, we provide a phasing formula
where we also deform the PN terms in the orbital energy of the binary.
This should be seen as a first step towards a more complete parametrized
phasing where we separate the conservative and dissipative contributions
to the phasing. 
A systematic revisit  of the problem
starting from the foundations of PN theory as applied to the compact
binary is needed to obtain a complete phasing formula parametrizing
uniquely the conservative and dissipative sectors in the phasing formula. We
postpone this for a follow up work.

The present results using nonspinning waveforms should be considered to
be a proof-of-principle demonstration, to be followed up with a more
realistic waveform that accounts for spin effects, effects of orbital
eccentricity and higher modes. The incorporation of the proposed test in the
framework of the effective one-body formalism~\cite{BuonD98} is also among
the future directions we plan to pursue. There are ongoing efforts to implement 
this method in the framework of {\tt LALInference}~\cite{lalinference} so 
that it can be applied to the compact binaries detected by advanced 
LIGO and Virgo detectors.

\section*{Acknowledgments} 
S.K. and K.G.A. thank B. Iyer, G. Faye, A. Ashtekar, G. Date, A. Ghosh and J. Hoque for
several useful discussions and  N. V. Krishnendu for cross-checking
some of the calculations reported here. We thank B. Iyer for useful comments and
suggestions on the manuscript as part of the internal review of the
LIGO and Virgo Collaborations, which has helped us improve the
presentation. K.G.A., A.G., S.K. and B.S.S. acknowledge the support
by the Indo-US Science and Technology Forum through the Indo-US {\em
Centre for the Exploration of Extreme Gravity}, grant
IUSSTF/JC-029/2016. A.G. and B.S.S. are supported in part by NSF grants
PHY-1836779, AST-1716394 and AST-1708146.  K.G.A. is partially support by a
grant from the Infosys Foundation. K.G.A. also acknowledges partial support by
the grant EMR/2016/005594. C.V.D.B. is supported by the research programme of the Netherlands Organisation for Scientific Research (NWO). Computing resources for this project were provided by the Pennsylvania State University.  This document has LIGO preprint number {\tt P1800274}.

\appendix*
\section{Frequency-domain phasing formula allowing for the deformation
of conservative dynamics} 

The binding energy parametrized at each PN order by four different constants $\{\alpha_0, \alpha_1,\alpha_2,\alpha_3\}$ used in the computation of parametrized GW phasing considering deviations in the conserved energy (mentioned in Sec.~\ref{sec2B}), is given by

\begin{eqnarray}\label{eq:EConsScaled}
E(v)&=&-\frac{1}{2} \nu v^2   \left[ \alpha_0 - 
\left (\frac{3}{4} + \frac{1}{12} \nu \right ) \alpha_1 v^2 - \left( \frac{27}{8} -\frac{19}{8} \nu  + \frac{1}{24}\nu^2 \right) \alpha_2 v^4\right.\nonumber\\
&&\left.- \left\{ \frac{675}{64}
- \left(\frac{34445}{576}-\frac{205 }{96}\pi^2\right)\nu 
+\frac{155 }{96}\nu^2
\right.\right.\nonumber\\
&&\left.\left.+\frac{35 }{5184}\nu^3
\right\} \alpha_3 v^6 \right]\,.
\end{eqnarray}
The resulting phase is quoted below,

\begin{widetext}\begin{eqnarray}
	\psi(f)&=&2\pi f t_c-\frac{\pi}{4}-\phi_c+\frac{3\alpha_0}{128v^5\mu_2^2\nu}\Bigg\{1+
	v^2\Bigg(\frac{2140}{189}-\frac{1100}{189}\nu-\frac{\alpha_1}{\alpha_0}\Bigg[\frac{10}{3}+\frac{10}{27}\nu\Bigg]+\hat{\mu}_3^2\Bigg[-\frac{6835}{2268}+\frac{6835}{567}\nu\Bigg]+\hat{\epsilon_2}^2\Bigg[-\frac{5}{81}+ \frac{20}{81}\nu\Bigg]\Bigg)
	\linechange
	-16\pi v^3+
	v^4\Bigg(\frac{295630}{1323}-\frac{267745}{2646} \nu +\frac{32240}{1323} \nu ^2 +\frac{\alpha_1}{\alpha_0}\Bigg[-\frac{535}{7}+\frac{1940}{63}\nu+\frac{275}{63}\nu^2\Bigg]+\frac{\alpha_2}{\alpha_0}\Bigg[-\frac{405}{4}+\frac{285}{4}\nu-\frac{5}{4}\nu^2\Bigg]
	\linechange
	+{\hat{\mu}_3}^2\Bigg[-\frac{104815}{3528}+\frac{8545 }{63} \nu-\frac{29630 }{441}\nu^2+\frac{\alpha_1}{\alpha_0}\Bigg(\frac{6835}{336}-\frac{34175}{432} \nu-\frac{6835}{756}\nu^2\Bigg)\Bigg]+{\hat{\mu}_3}^2\hat{\epsilon_2}^2\Bigg[\frac{6835}{9072}
	-\frac{6835}{1134} \nu +\frac{6835 \nu ^2}{567}\Bigg]
	\linechange
	+\hat{\mu}_3^4\Bigg[\frac{9343445}{508032}-\frac{9343445}{63504} \nu +\frac{9343445}{31752} \nu ^2\Bigg]+\hat{\mu}_4^2\Bigg[-\frac{89650}{3969}+\frac{179300}{1323} \nu  - \frac{89650}{441}\nu^2\Bigg]
	+\hat{\epsilon_2}^2 \Bigg[-\frac{1885}{756}+\frac{695}{63}\nu 
	\linechange
	-\frac{800}{189} \nu ^2
	+\frac{\alpha_1}{\alpha_0}\Bigg(\frac{5}{12}-\frac{175}{108}\nu -\frac{5}{27}\nu ^2\Bigg)\Bigg]
	+\hat{\epsilon_2}^4\Bigg[\frac{5}{648}-\frac{5 }{81} \nu+\frac{10}{81} \nu ^2
	\Bigg]
	+\hat{\epsilon_3}^2\Bigg[-\frac{50}{63}+\frac{100}{21} \nu-\frac{50}{7} \nu ^2\Bigg]
	\Bigg)
	\linechange
	+\pi v^5 \Big(3 \log\left[\frac{v}{v_{\rm LSO}}\right]+1\Big) \Bigg(\frac{80}{189} \Big[214-131 \nu \Big]-\frac{80\alpha_1}{27\alpha_0}\Big[9+\nu\Big]
	-\frac{9115}{756}\hat{\mu}_3^2\Big[ 1-4 \nu\Big]- \frac{20}{27} \hat{\epsilon_2}^2\Big[ 1-4 \nu\Big]
	\Bigg)
	\linechange
	+v^6\Bigg(\frac{36847016}{509355}-\frac{640}{3}\pi^2-\frac{6848}{21}\gamma_E -\frac{6848}{21} \log[4v]+\Bigg[\frac{28398155}{67914}+\frac{205}{12}\pi^2\Bigg]\nu -\frac{563225}{3773}\nu^2+ \frac{3928700}{305613}\nu^3
	\linechange
	+\frac{\alpha_1}{\alpha_0}\Bigg[\frac{295630}{441}-\frac{1818445}{7938}\nu +\frac{312575}{7938}\nu^2+ \frac{32240}{3969}\nu^3\Bigg]+\frac{\alpha_2}{\alpha_0}\Bigg[\frac{14445}{14}-\frac{8795}{7}\nu +\frac{8105}{21}\nu^2 - \frac{275}{42}\nu^3\Bigg]
    \linechange
    +\frac{\alpha_3}{\alpha_0}\Bigg[\frac{3375}{4}+\Bigg(-\frac{172225}{36}+\frac{1025}{6}\pi^2\Bigg)\nu +\frac{775}{6}\nu^2+ \frac{175}{324}\nu^3\Bigg]
	+\hat{\mu}_3^2\Bigg[
	\frac{732782515}{3667356}-\frac{1061322545}{1222452} \nu
	+\frac{1027073335}{3667356} \nu^2
	\linechange
	-\frac{15723035}{916839} \nu^3 + \frac{\alpha_1}{\alpha_0}\Bigg(-
	\frac{104815}{1176}+\frac{4201865}{10584} \nu
	-\frac{206855}{1323} \nu^2-\frac{29630}{1323}\nu^3\Bigg) 
	+ \frac{\alpha_2}{\alpha_0}\Bigg(-
	\frac{61515}{224}+\frac{868045}{672} \nu 
	-\frac{1565215}{2016} \nu^2
	\linechange
	+\frac{6835}{504}\nu^3\Bigg)
	\Bigg]
	+\hat{\mu}_3^2\hat{\epsilon_2}^2 \Bigg[-\frac{1742995}{190512}+\frac{1045805}{13608}\nu
	-\frac{2091650}{11907}\nu ^2+\frac{697310}{11907}\nu ^3 +\frac{\alpha_1}{\alpha_0}\Bigg(\frac{6835}{3024}-\frac{485285}{27216}\nu
	+\frac{116195}{3402}\nu ^2+\frac{6835}{1701}\nu ^3\Bigg)
	\Bigg]
	\linechange
	+\hat{\mu}_3^2\hat{\epsilon_2}^4\Bigg[\frac{6835}{108864}-\frac{6835
	}{9072} \nu+\frac{6835}{2268} \nu^2-\frac{6835}{1701} \nu^3
	\Bigg]
	+\hat{\mu}_3^2\hat{\epsilon_3}^2\Bigg[-\frac{34175}{7938
	}+\frac{170875}{3969} \nu 
	-\frac{375925}{2646}\nu ^2+\frac{68350}{441} \nu ^3
	\Bigg]
	\linechange
	+\hat{\mu}_3^2\hat{\mu}_4^2\Bigg[-\frac{61275775}{500094}+\frac{306378875 
	}{250047}\nu-\frac{674033525}{166698} \nu^2 +\frac{122551550}{27783} \nu^3
	\Bigg]
	+\hat{\mu}_3^4\Bigg[\frac{140055985}{5334336}-\frac{1148286835}{5334336} \nu+\frac{307950925}{666792}\nu^2
	\linechange
	-\frac{27838955}{333396} \nu^3 +\frac{\alpha_1}{\alpha_0}\Bigg(\frac{9343445}{169344}-\frac{663384595}{1524096} \nu
	+\frac{158838565}{190512}\nu^2
	+\frac{9343445}{95256} \nu^3\Bigg)\Bigg]
	+\hat{\mu}_3^4\hat{\epsilon_2}^2\Bigg[\frac{9343445}{3048192} -\frac{9343445}{254016}\nu
	\linechange
	+\frac{9343445}{63504} \nu^2-\frac{9343445}{47628} \nu^3
	\Bigg]
	+ \hat{\mu}_3^6\Bigg[\frac{12772489315}{256048128}-\frac{12772489315 }{21337344} \nu+\frac{12772489315}{5334336} \nu^2-\frac{12772489315}{4000752} \nu^3\Bigg]
	\linechange
	+\hat{\mu}_4^2\Bigg[-\frac{24426860}{916839}+\frac{62508560}{305613} \nu
	-\frac{12980600}{33957} \nu^2+\frac{286700}{11319} \nu ^3 +\frac{\alpha_1}{\alpha_0}\Bigg(-\frac{89650}{1323}+\frac{4751450}{11907} \nu-\frac{2241250}{3969} \nu^2-\frac{89650}{1323} \nu ^3\Bigg) \Bigg]
	\linechange
	+ \hat{\mu}_4^2\hat{\epsilon_2}^2\Bigg[-\frac{89650}{35721}+\frac{896500}{35721}\nu
	-\frac{986150}{11907}\nu^2+\frac{358600 \nu ^3}{3969}
	\Bigg]
	+\hat{\mu}_5^2\Bigg[ \frac{1002569}{12474}-\frac{4010276}{6237} \nu+\frac{10025690}{6237} \nu^2-\frac{8020552 }{6237}\nu^3\Bigg]
	\linechange
	+\hat{\epsilon_2}^2\Bigg[\frac{6134935}{190512}-\frac{2353285}{15876}\nu+\frac{550075}{6804}\nu^2-\frac{150845}{11907} \nu^3 +\frac{\alpha_1}{\alpha_0}\Bigg(-\frac{1885}{252}+\frac{73175}{2268}\nu-\frac{1705}{189}\nu^2-\frac{800}{567} \nu^3 \Bigg) 
	\linechange
	+\frac{\alpha_2}{\alpha_0}\Bigg(-\frac{45}{8}+\frac{635}{24}\nu-\frac{1145}{72}\nu^2+\frac{5}{18} \nu^3 \Bigg)\Bigg]
	+\hat{\epsilon_2}^2\hat{\epsilon_3}^2\Bigg[-\frac{50}{567}+\frac{500}{567} \nu -\frac{550}{189} \nu^2+\frac{200}{63} \nu ^3\Bigg ]
	+\hat{\epsilon_2}^4\Bigg[-\frac{25}{126}+\frac{3775}{2268} \nu-\frac{2150}{567}\nu^2
	\linechange
	+\frac{100}{81} \nu ^3 + \frac{\alpha_1}{\alpha_0}\Bigg(\frac{5}{216}-\frac{355}{1944} \nu+\frac{85}{243}\nu^2+\frac{10}{243}\nu^3\Bigg) \Bigg]
	+\hat{\epsilon_2}^6\Bigg[\frac{5}{11664}-\frac{5}{972} \nu +\frac{5}{243} \nu ^2 -\frac{20}{729} \nu^3\Bigg]
	+\hat{\epsilon_3}^2\Bigg[\frac{37180}{3969}-\frac{235640}{3969} \nu
	\linechange
	+\frac{420200}{3969} \nu^2- \frac{47900}{1323}\nu^3 + \frac{\alpha_1}{\alpha_0}
	\Bigg(-\frac{50}{21}+\frac{2650}{189} \nu -\frac{1250}{63} \nu^2- \frac{50}{21}\nu^3 \Bigg) \Bigg]
	+\hat{\epsilon_4}^2\Bigg[\frac{5741}{1764}-\frac{11482}{441} \nu+\frac{28705 }{441}\nu ^2 -\frac{22964}{441}\nu ^3\Bigg]\Bigg)
	\linechange
	+\pi v^7\Bigg(\frac{2365040}{1323}-\frac{1300930}{1323}\nu+\frac{400930}{1323}\nu^2 +\frac{\alpha_1}{\alpha_0}\Bigg[-\frac{4280}{7}+\frac{19300}{63}\nu+\frac{2620}{63}\nu^2
	\Bigg]+\frac{\alpha_2}{\alpha_0}\Big(-810+570\nu-10\nu^2\Big)
	\linechange
	+\hat{\mu}_3^2\Bigg[-\frac{69905}{588}+\frac{191495}{336}\nu-\frac{73995}{196}\nu^2 +\frac{\alpha_1}{\alpha_0}\Bigg(\frac{9115}{112}-\frac{45575}{144}\nu-\frac{9115}{252}\nu^2\Bigg)\Bigg]
	+\hat{\mu}_3^2\hat{\epsilon_2}^2\Bigg[\frac{54685}{9072}
	-\frac{54685}{1134}  \nu+\frac{54685 }{567}  \nu^2\Bigg]
	\linechange
	+\hat{\mu}_3^4\Bigg[\frac{6835}{254016}-\frac{6835}{31752}  \nu+\frac{6835}{15876}\nu^2\Bigg]
	+\hat{\mu}_4^2\Bigg[-\frac{400}{3969}+\frac{800}{1323} \nu -\frac{400}{441}\nu ^2\Bigg]
	+\hat{\epsilon_2}^2\Bigg[-\frac{1885}{63}+\frac{2815}{21}\nu -\frac{3620}{63}\nu ^2 	
	\linechange
	+\frac{\alpha_1}{\alpha_0}\Bigg(5-\frac{175}{9}\nu-\frac{20}{9}\nu^2\Bigg)\Bigg]
	+\hat{\epsilon_3}^2\Bigg[-\frac{400 }{63}+\frac{800}{21} \nu -\frac{400}{7}\nu ^2
	\Bigg]
	+\hat{\epsilon_2}^4\Bigg[\frac{10}{81}-\frac{80}{81} \nu +\frac{160 }{81}\nu ^2\Bigg]
	\Bigg)\Bigg\}\,.\label{eq:psicons}
	\end{eqnarray}
\end{widetext}

The GW phasing for compact binaries can be represented by various PN
approximants depending on the different ways in which they treat the energy and flux
functions. We refer the reader to Refs.~\cite{DIS02, BIOPS09}
for a detailed discussion of these various approximants. We provide the input functions required for the computation of the
phasing for TaylorT2, TaylorT3 and TaylorT4 in a {\tt Mathematica} file (supl-Multipole.m) which serves the Supplemental Material to this paper. 
We closely follow the notations of Ref.~\cite{BIOPS09} in this file.

\bibliographystyle{apsrev}
\bibliography{./Ref_list}

\end{document}